\documentclass[conference]{IEEEtran}

\usepackage{cite}
\usepackage{tikz}
\usepackage{tabularx}
\usepackage{enumitem}
\usepackage{dblfloatfix}
\usepackage{url}
\usepackage{hyperref}
\usepackage{array,multirow,graphicx}
\usepackage{latexsym}
\usepackage{bm}
\usepackage{mathtools}
\usepackage{etoolbox}
\usepackage[font=small,skip=2pt]{caption}
\usepackage{pgfplots,pgfplotstable}
\usepackage{booktabs}
\usepackage{cleveref}
\usepackage{multicol}
\usepackage{microtype}
\usepackage{color}
\usepackage[boxed]{algorithm2e}
\usepackage[noend]{algpseudocode}
\usepackage[compact]{titlesec}
\algnewcommand{\IfThenElse}[3]{% \IfThenElse{<if>}{<then>}{<else>}
  \State \algorithmicif\ #1\ \algorithmicthen\ #2\ \algorithmicelse\ #3}
\makeatletter
\newcommand{\removelatexerror}{\let\@latex@error\@gobble}
\makeatother
\usepackage{subfigure}
\usepackage{letltxmacro}

\newcommand{\mytilde}{\raise.17ex\hbox{$\scriptstyle\mathtt{\sim}$}}

\usepackage{amsthm}

\usepackage{amsmath}

\SetKwInput{KwInput}{Input}                % Set the Input
\SetKwInput{KwOutput}{Output}  

\usepackage{tikz}
\usepackage{capt-of}
\usepackage{amssymb}% http://ctan.org/pkg/amssymb
\usepackage{pifont}% http://ctan.org/pkg/pifont

\newcommand\blfootnote[1]{%
  \begingroup
  \renewcommand\thefootnote{}\footnote{#1}%
  \addtocounter{footnote}{-1}%
  \endgroup
}
\begin{document}

\title{Modeling Silicon-Photonic Neural Networks under Uncertainties}
\author{\IEEEauthorblockN{Sanmitra Banerjee\IEEEauthorrefmark{2}, Mahdi Nikdast\IEEEauthorrefmark{3}, and Krishnendu Chakrabarty\IEEEauthorrefmark{2}}\\ \vspace{-0.3cm}
\IEEEauthorblockA{\IEEEauthorrefmark{2}Department of Electrical and Computer Engineering, Duke University, Durham, NC 27708, USA}
\IEEEauthorblockA{\IEEEauthorrefmark{3}Department of Electrical and Computer Engineering, Colorado State University, Fort Collins, CO 80523, USA}
%\vspace{-1em}
}

\maketitle
\bstctlcite{IEEEexample:BSTcontrol}
\pagenumbering{gobble}

% Page numbering ... 
\thispagestyle{plain}
\pagestyle{plain}
\linespread{0.96}

%[==============================================
% Paper contents

\begin{abstract}
Silicon-photonic neural networks (SPNNs) offer substantial improvements in computing speed and energy efficiency compared to their digital electronic counterparts. However, the energy efficiency and accuracy of SPNNs are highly impacted by uncertainties that arise from fabrication-process and thermal variations. In this paper, we present the first comprehensive and hierarchical study on the impact of random uncertainties on the classification accuracy of a Mach--Zehnder Interferometer (MZI)-based SPNN. We show that such impact can vary based on both the location and characteristics (e.g., tuned phase angles) of a non-ideal silicon-photonic device. Simulation results show that in an SPNN with two hidden layers and 1374 tunable-thermal-phase shifters, random uncertainties even in mature fabrication processes can lead to a catastrophic 70\% accuracy loss. 
\end{abstract}

\section{Introduction}%\vspace{-0.05in}
\blfootnote{\hspace{-1.0em}-----------------------------------------------------\\ This  research was supported in part by the National Science Foundation (NSF) under grant CCF-2006788.}
In deep neural networks (DNNs), matrix multiplication is known to be the most
time- and energy-intensive operation. Silicon-photonic neural networks (SPNNs) employ photonic components to optimize matrix multiplication with ultra-high speed and ultra-low energy consumption \cite{cheng2020silicon}. The linear multipliers are represented using two unitary multipliers and a diagonal matrix, which are obtained using singular value decomposition (SVD). The multipliers and the diagonal matrix can be realized using a network of interconnected Mach--Zehnder interferometers (MZIs) \cite{clements2016optimal}. In the absence of optical crosstalk, the complexity of matrix-vector multiplication can be reduced from $O(N^2)$ to $O(1)$ \cite{cheng2020silicon}. However, there exist several roadblocks in the further advancement of SPNNs; these include the optical loss associated with MZI networks \cite{clements2016optimal,reck1994experimental}, additional computation needed for mapping the trained weights to the parameters (i.e., phase angles) in MZI arrays \cite{clements2016optimal}, and the finite-encoding precision on phase settings \cite{cheng2020silicon}.  

In this paper, we present the first comprehensive analysis
of the impact of uncertainties due to fabrication-process variations (FPVs) and thermal crosstalk in SPNNs. Perturbations in specific MZIs,
depending on their position and tuned phase angles, can be
catastrophic in nature. Therefore, identifying such components
during the design time is necessary for improving the yield. To address this requirement, we develop a framework to identify critical components in SPNNs where random uncertainties lead to severe performance degradation in the network. Significant degradation in SPNN performance (70\% loss in inferencing accuracy) is observed considering typical uncertainties---reported in prior work \cite{flamini2017benchmarking}---in the MZIs. 

\begin{figure}[t]
  \centering
  \includegraphics[scale=0.385]{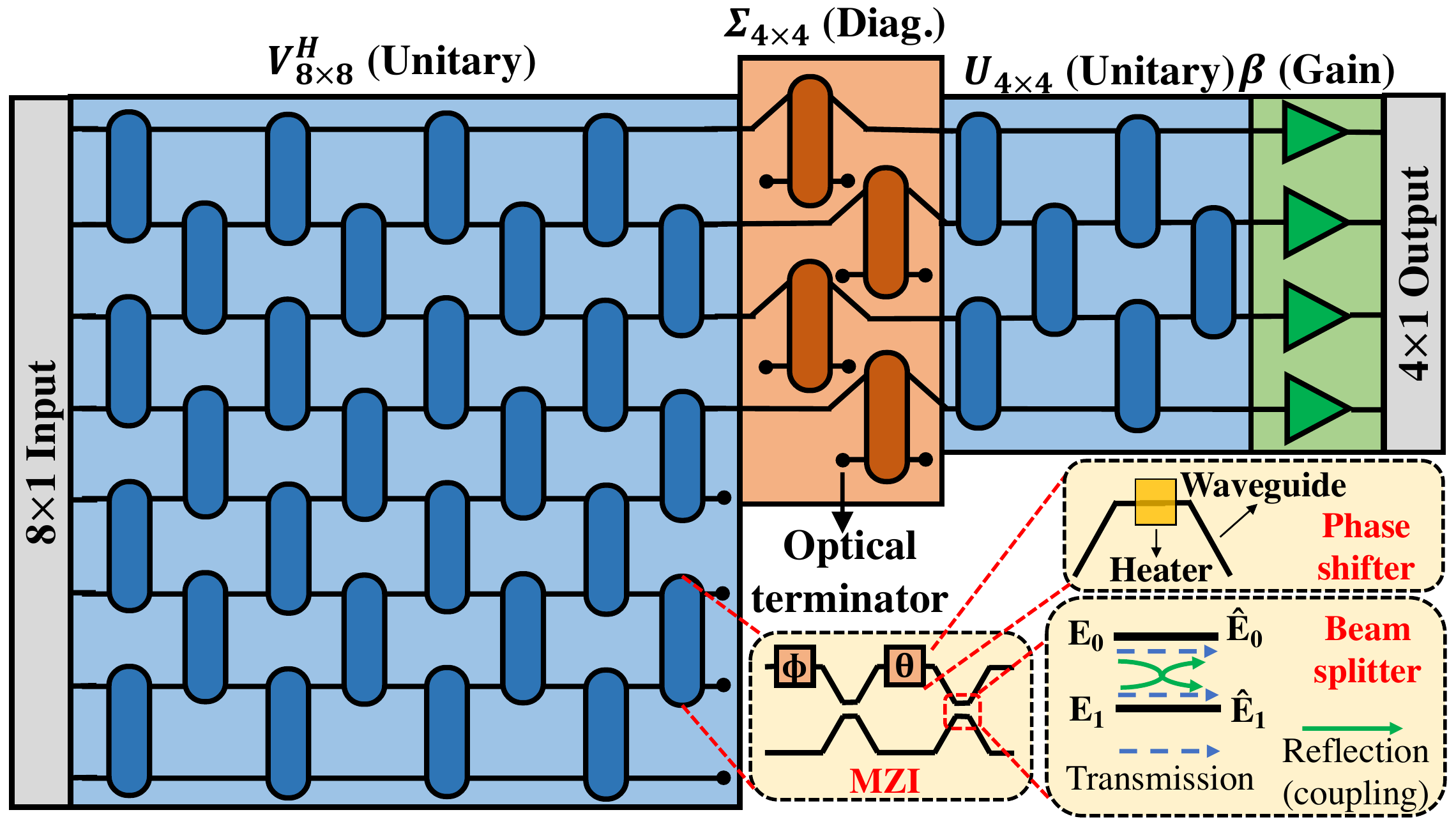}
  \caption{SPNN linear-layer representation using MZI arrays. An 8$\times$4 linear layer is represented in this example. Bottom: An MZI structure.}
  \vspace{-1.8em}
  \label{placeholder}
\end{figure}

\section{Background and Motivation}

\subsection{Mach--Zehnder Interferometer (MZI)}
As shown in Fig. 1, a typical MZI consists of two tunable phase shifters (PhS, $\phi$ and $\theta$) on the upper arm and two 50:50 beam splitters (BeS). The PhS are used to apply configurable phase shifts and obtain varying degrees of interference between the input optical signals. They can be implemented using thermal microheaters, where the refractive index of the underlying waveguide changes with temperature (i.e., thermo-optic effect), altering the phase of the optical signal traversing the waveguide. Moreover, 2$\times$2 BeS can be designed using directional couplers, where a fraction (defined by transmittance) of the optical signal at an input port is transmitted to an output port, and the remaining (defined by the reflectance) is coupled to the other output port with a phase shift of $\frac{\pi}{2}$. For symmetric 50:50 BeS, both transmittance and reflectance coefficients are $\frac{1}{\sqrt{2}}$. As a result, the transfer matrix for an MZI with two PhS and two 50:50 BeS (see Fig. \ref{placeholder}) can be defined as \cite{fang2019design}:
\begin{equation}
    \begin{split}
       &T_{MZI}(\theta, \phi)=U_{BeS}\cdot U_{PhS}(\theta)\cdot U_{BeS}\cdot U_{PhS}(\phi)\\
       &=\begin{pmatrix}
       T_{11} & T_{12} \\
        T_{21} & T_{22}
       \end{pmatrix} =  \begin{pmatrix}
        \frac{e^{i\phi}}{2}(e^{i\theta}-1) & \frac{i}{2}(e^{i\theta}+1) \\
        \frac{ie^{i\phi}}{2}(e^{i\theta}+1) & -\frac{1}{2}(e^{i\theta}-1) 
        \end{pmatrix} 
    \end{split},
\end{equation}
where $U_{BeS}$ ($U_{PhS}$) is the BeS (PhS) transfer matrix. \par

\subsection{Design of MZI-based SPNNs}
Fully connected layers can be represented mathematically as matrix-vector multiplication followed by an activation function. Consider a layer $L_i$ with $n_i$ neurons fully connected to the previous layer $L_{i-1}$ with $n_{i-1}$ neurons. The output vector at $L_{i}$ is then given by \mbox{$O_{i}^{n_{i}\times 1}=f_{i}(M_{i}^{n_{i}\times n_{i-1}}O_{i-1}^{n_{i-1}\times 1})$}. Note that $f_{i}$ and $M_{i}$ are the non-linear activation function and weight matrix associated with layer $L_{i}$, respectively. In SPNNs, the linear multiplication with the weight matrix (i.e., $M_i$) is often implemented using arrays of configurable MZIs. Using SVD and considering Fig. 1, we have $M_{i}=U_{i}\Sigma_{i}V_{i}^H$, where $U_{i}$ and $V_{i}$ are unitary matrices with dimensions $n_{i}\times n_{i}$ and $n_{i-1}\times n_{i-1}$, respectively. Moreover, $V_{i}^H$ denotes the Hermitian transpose of $V_{i}$ and $\Sigma_{i}$ is a diagonal matrix consisting of the eigenvalues of $M_{i}$.
 
Given a weight matrix $M_{i}=U_{i}\Sigma_{i}V_{i}^H$, we use the Clements design \cite{clements2016optimal} to represent the unitary matrices $U_{i}$ and $V_{i}^H$. The diagonal matrix $\Sigma_{i}$ can be represented using a similar MZI array where one input and one output of each MZI are terminated (see Fig. \ref{placeholder}). A global optical amplification is necessary on each output to represent arbitrary diagonal matrices \cite{connelly2007semiconductor}. This scaling factor is realized using layer $\beta$, as shown in Fig. \ref{placeholder}.\par

\subsection{Related Work on Component Imprecision in SPNNs}
Deviations in the phase angles in PhS and the splitting ratios in BeS---due to inevitable FPVs and thermal crosstalk---have a severe impact on MZI performance in SPNNs \cite{lu2017performance}. The use of thermal actuators to compensate for phase errors leads to induced mutual thermal crosstalk between neighboring waveguides \cite{milanizadeh2019canceling}. A method to counter the impact of both FPVs and thermal effects using a modified cost function during training and post-fabrication hardware calibration is presented in \cite{ying2020variation}. However, this method only focuses on uncertainties in the phase angles, ignoring the considerable impact of inevitable errors in BeS. Moreover, the required hardware calibration necessitates the tuning of each MZI in the network, and this step becomes increasingly complex as the network scales up. The modified training method also results in accuracy loss.\par

Here, we model the impact of random and non-uniform uncertainties in both phase angles and beam-splitting ratios in MZIs in SPNNs. We also show that the impact of uncertainties depends both on the position and parameter values of the affected MZIs. Therefore, some random variations in some MZIs can be more critical than others. Our entire analysis can be performed prior to fabrication and after software training.

\section{Uncertainties in SPNNs: A Hierarchical Study}
In this section, we systematically analyze the impact of uncertainties on the performance of SPNNs in a hierarchical fashion at the component-level (PhS and BeS), device-level (MZIs), layer-level (MZI array), and system-level (SPNN).  

\subsection{Component-Level: Phase Shifters and Beam Splitters}
The temperature-dependent phase change in a thermo-optic PhS is given by $\Delta \phi=\left(\frac{2\pi l}{\lambda_0}\right)\cdot \left(\frac{dn}{dT}\right)\cdot \Delta T$, where $l$ is the length of the phase shifter and $\lambda_0$ is the optical wavelength \cite{jacques2019optimization}. Also, $\frac{dn}{dT}\approx$ 1.8$\times10^{-4} K^{-1}$ is the thermo-optic coefficient of silicon at $\lambda_{0}=$~1550~nm and temperature $T=$~300~$K$ \cite{walls2007quantum}, and $\Delta T$ is the temperature change.

During \textit{in-situ} training of SPNNs, the phase angles at PhS are applied using thermal actuators (i.e., microheaters). Mutual thermal crosstalk among neighboring actuated waveguides, which are placed in proximity in SPNNs (see Fig. \ref{placeholder}), affects the efficiency of the tuning and bias-control mechanism, imposing phase-angle errors.  
Furthermore, FPVs can change $l$ (see $\Delta\phi$), hence impacting the efficiency of PhS. Due to random perturbations in the phase angles ($\theta$ and $\phi$) in (1), $T_{MZI}$ will deviate from its intended form, resulting in faulty matrix multiplication and a reduction in SPNN inferencing accuracy. \par

Considering the classical, lossless 2$\times$2 beam-splitter schematic shown in Fig. \ref{placeholder}, the electric fields at the output $\tilde{E}_{0/1}$ can be attributed to the transmitted electric-field component $E_0$ and the reflected electric-field component $E_1$ based on \cite{fang2019design}:
\begin{equation}
    \begin{pmatrix}
        \tilde{E_0} \\
        \tilde{E_1} 
        \end{pmatrix}
        = \begin{pmatrix}
        r_{00} & it_{10} \\
        it_{01} & r_{11} 
        \end{pmatrix}
        \begin{pmatrix}
        E_0 \\
        E_1
        \end{pmatrix}.
\end{equation}

Here, $r$ and $t$ represent the reflectance and transmittance associated with each path, respectively. Note that $r_{00}^2+t_{01}^2=$~1 and $r_{11}^2+t_{10}^2=$~1. For symmetric BeS, $r_{00}=r_{11}=r$ and $t_{01}=t_{10}=t$. Additionally, for ideal 50:50 BeS, $r=t=\frac{1}{\sqrt{2}}$. However, under random FPVs, $r$ and $t$ will deviate from $\frac{1}{\sqrt{2}}$; this results in unbalanced and imperfect BeS \cite{liu2016compensation, Mahdi_JLT}. Unlike PhS, BeS are passive devices and once fabricated, we cannot actively change their $r$ and $t$ values during SPNN training.\par

Prior studies have shown an error of $\mytilde$0.21 radian in the tuned phase angles in PhS for mature fabrication processes \cite{flamini2017benchmarking}. This corresponds to $\frac{0.21}{2\pi}\times$100 $\approx$~3.34\% of the range of phase angles. Taking this into consideration, we perturb $\theta$ and $\phi$ using a Gaussian distribution with mean ($\mu$) set to their nominal tuned values (obtained from training) and multiple values of standard deviation in the range \mbox{$0.005\cdot 2\pi\leq\sigma\leq 0.15\cdot 2\pi$}. While a deviation of 1--2\% is typically expected in the $r$ and $t$ parameters in BeS \cite{flamini2017benchmarking}, we vary them using a similar distribution as PhS---Gaussian with $\mu=\frac{1}{\sqrt{2}}$ and $0.005\cdot \frac{1}{\sqrt{2}}\leq\sigma\leq 0.15\cdot \frac{1}{\sqrt{2}}$---for a fair comparison of their impact on accuracy. In the rest of the paper, we use $\sigma_{PhS}$ to refer to $\frac{\sigma}{2\pi}$ for PhS, and $\sigma_{BeS}$ to refer to $\sqrt{2}\sigma$ for BeS.

\subsection{Device-Level: MZIs}
Variations in $\theta$ ($\Delta\theta$) and $\phi$ ($\Delta \phi$) phase angles in PhS can result in deviations in the MZI transfer matrix, $T_{MZI}$, defined in (1). Such deviations can be defined as:
\begin{equation}
\begin{split}
      & \Delta T_{MZI}(\theta, \phi) = \frac{\partial T_{MZI}(\theta, \phi)}{\partial \theta}\Delta \theta + \frac{\partial T_{MZI}(\theta, \phi)}{\partial \phi}\Delta \phi \\
    & =\begin{pmatrix}
    \frac{ie^{i(\phi+\theta)}}{2} & -\frac{e^{i\theta}}{2} \\
    -\frac{e^{i(\phi+\theta)}}{2} & -\frac{ie^{i\theta}}{2}
    \end{pmatrix}
    \Delta\theta + 
    \begin{pmatrix}
    \frac{ie^{i\phi}}{2}(e^{i\theta}-1) & 0 \\
    -\frac{e^{i\phi}}{2}(e^{i\theta}+1) & 0
    \end{pmatrix}
    \Delta\phi.  
\end{split}
\end{equation}
Let the relative changes in $\theta$ and $\phi$ be $K_{\theta}=\frac{\Delta\theta}{\theta}$~and \mbox{$K_{\phi}=\frac{\Delta\phi}{\phi}$}, respectively. We assume $K_{\theta}=K_{\phi}=K$ as the two PhS, corresponding to $\theta$ and $\phi$, are in proximity (see Fig. 1). Note that this assumption is made to simplify the analyses only in this subsection. In all subsequent analyses, independent variations are considered in $\theta$ and $\phi$. Thus, from (3), we have: 
\begin{equation}
    \Delta T_{MZI}(\theta, \phi)=K
    \begin{pmatrix}
    (\theta+\phi)\frac{ie^{i(\theta+\phi)}}{2}-\phi\frac{ie^{i\phi}}{2} & -\theta\frac{e^{i\theta}}{2} \\
    -(\theta+\phi)\frac{e^{i(\theta+\phi)}}{2}-\phi\frac{e^{i\phi}}{2} & -\theta\frac{ie^{i\theta}}{2}
    \end{pmatrix}.
\end{equation}

\begin{figure}[t]
\centering
\subfigure[$T_{11}$]{
\includegraphics[scale=0.2]{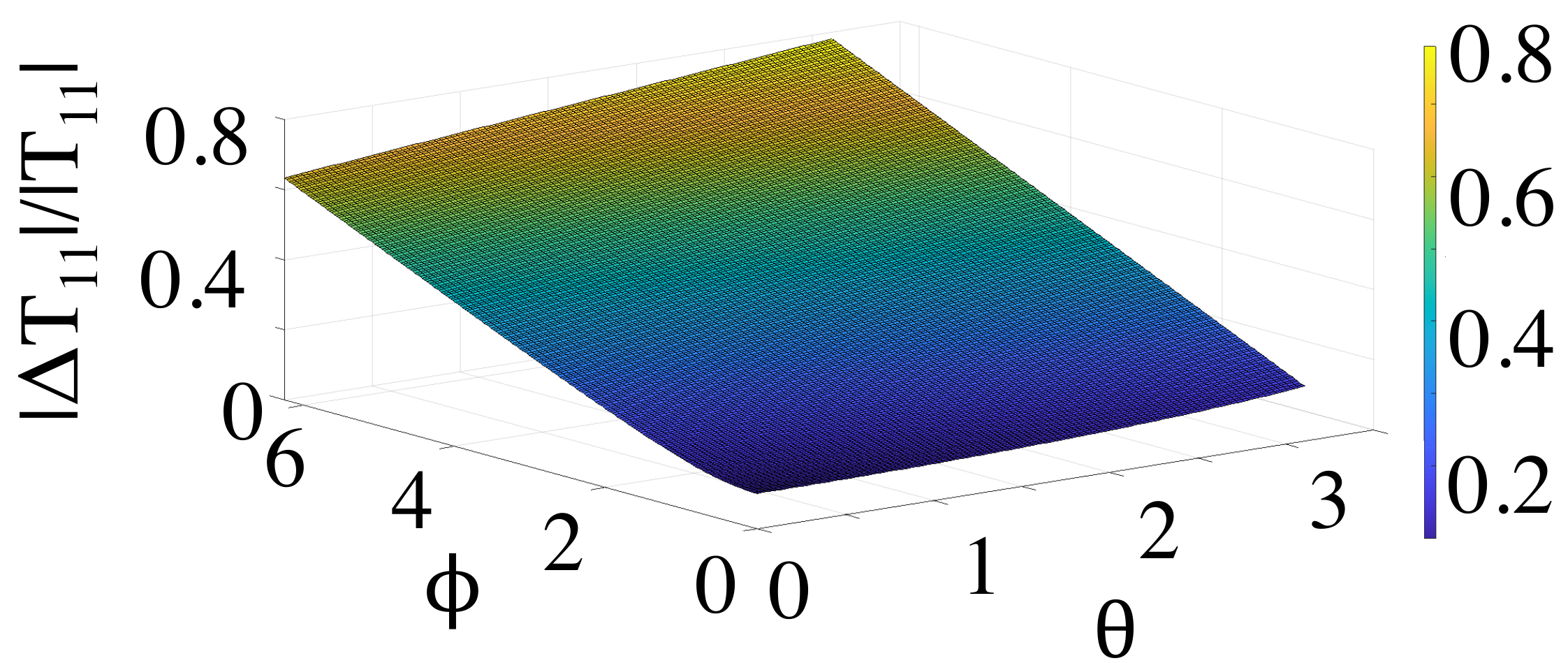}\label{fab}}\hspace{-0.3em}%
\subfigure[$T_{12}$]{
\includegraphics[scale=0.2]{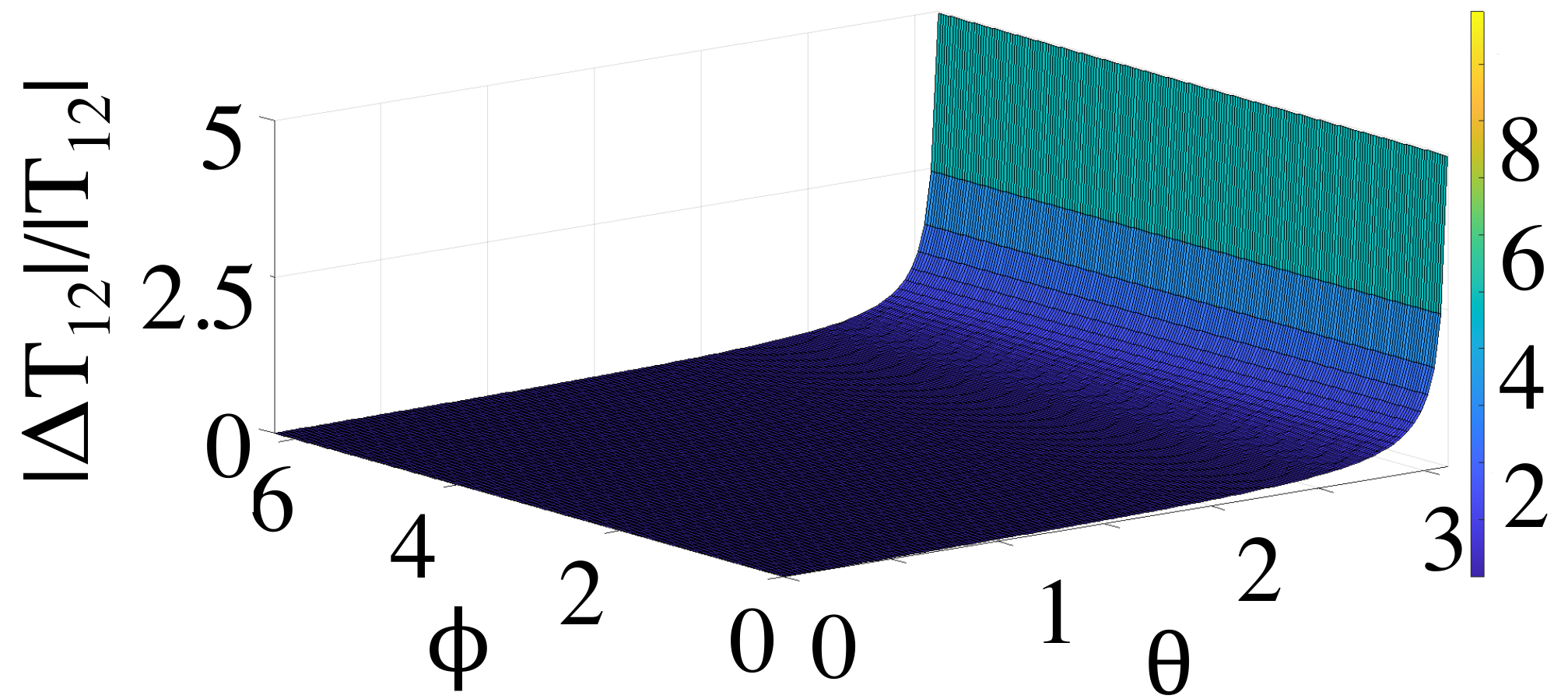}\label{slps}}%
\newline
\subfigure[$T_{21}$]{
\includegraphics[scale=0.2]{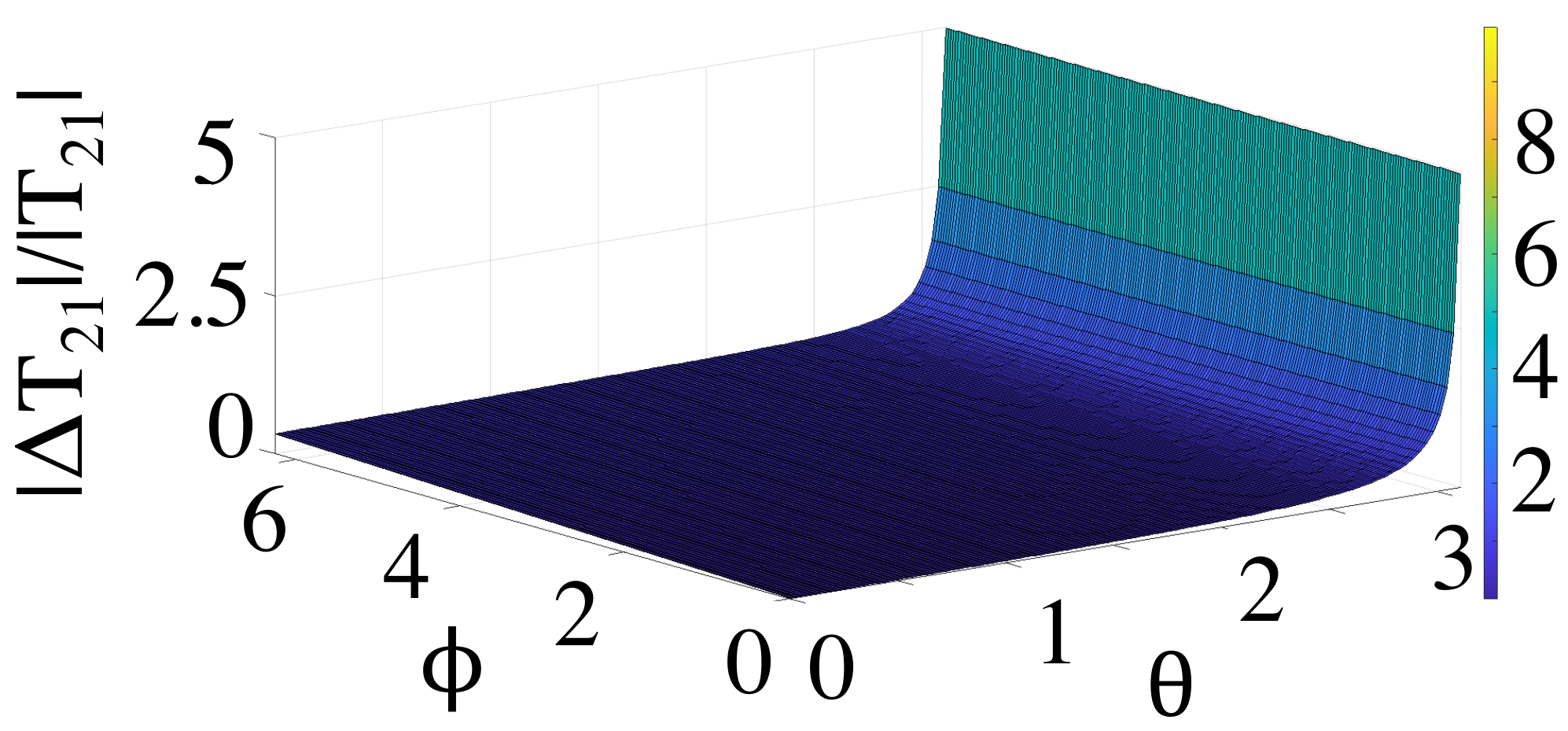}\label{total_shift}}\hspace{-0.3em}%
\subfigure[$T_{22}$]{
\includegraphics[scale=0.2]{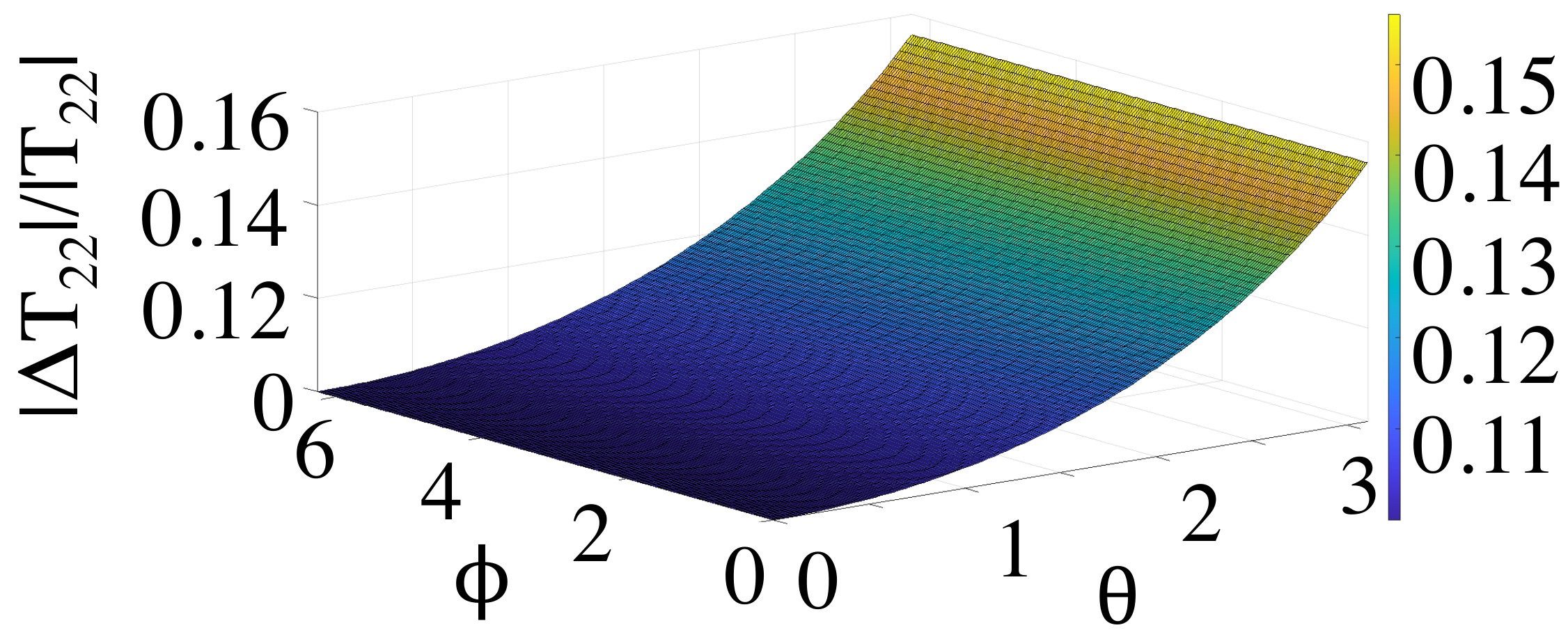}\label{fab}}%
\vspace{-0.05in}
\caption{Magnitude of variation in the absolute value of elements in $T_{MZI}$ (see (1)) relative to the modulus of their nominal values.}\label{Tmn_results}
%\squeezeup
\vspace{-0.25in}
\end{figure}

Using (1) and (4), Fig. \ref{Tmn_results} shows the magnitude of deviation for each of the four elements in $T_{MZI}$ relative to the modulus of their nominal values for different values of $\theta$ and $\phi$ with $K=$~0.05. We find that the relative deviation increases monotonically as $\theta$ and $\phi$ increase. This indicates that MZIs with higher values of tuned phase angles are more susceptible to uncertainties. \par

The proposed $T_{MZI}$ model in (1) assumes ideal 50:50 BeS with $r_{00}=r_{11}=t_{01}=t_{10}=\frac{1}{\sqrt{2}}$. However, under uncertainties in BeS, this model changes to:
\begin{equation}
    T_{MZI}(\theta, \phi)=\begin{pmatrix}
    rr'e^{i(\theta+\phi)}-tt'e^{i\phi} & ir'te^{i\theta}+it'r \\
    it're^{i(\theta+\phi)}+itr'e^{i\phi} & -tt'e^{i\theta}+rr'
    \end{pmatrix}.
\end{equation}
Here, $r~(t)$ and $r'~(t')$ are the reflectances (transmittances) for the first and the second beam splitter, respectively (see Fig. \ref{placeholder}). 

\subsection{Layer-Level: MZI Array}
Under uncertainties, $T_{MZI}$ deviates, and consequently, the matrix represented by the array can vary from the intended unitary matrix. We use the relative-variation distance (RVD) as a figure-of-merit to quantify the difference between the intended unitary matrix ($\tilde{U}$) and the deviated unitary matrix ($U$). This is given by \mbox{$RVD(U, \tilde{U})=\frac{\sum\limits_{m}\sum\limits_{n}\left|U_{m,n}-\tilde{U}_{m,n}\right|}{\left|\tilde{U}_{m,n}\right|}$}. \par

\begin{figure}[t]
  \centering
  \includegraphics[scale=0.42]{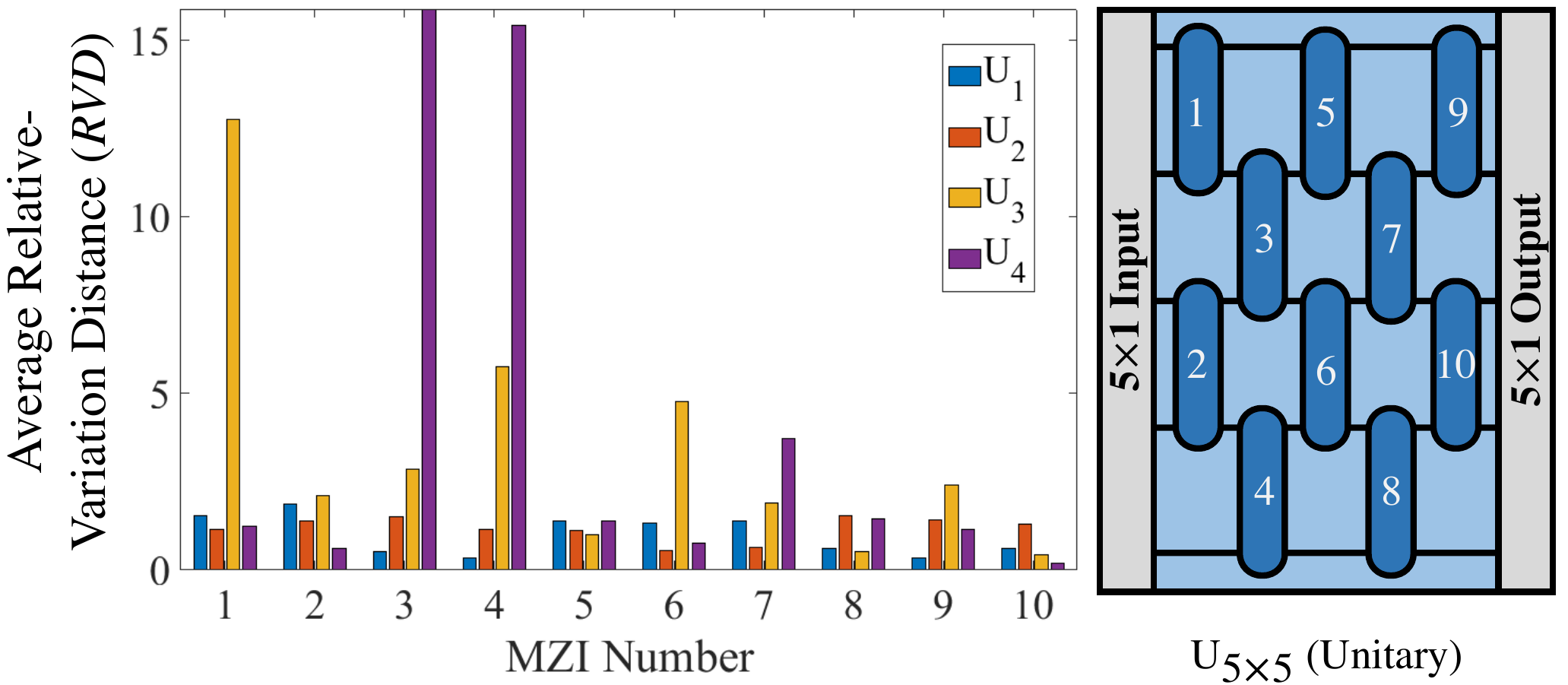}
  \caption{Average $RVD$ (left) for four random 5$\times$5 unitary matrices with one MZI under variations at a time. Right: An MZI array (including the MZI numbers) to represent any 5$\times$5 unitary matrix (see Fig. 1).}
  \vspace{-1.8em}
  \label{U1234}
\end{figure}

Different elements of the unitary transfer matrix are affected by different subsets of MZIs in the array. Therefore, variations in each MZI will have a unique impact on the overall $RVD$ defined above. This is indeed the case as is shown in Fig. \ref{U1234}. We consider four randomly generated 5$\times$5 unitary matrices with random perturbations in the PhS and BeS. For each matrix, we introduce variations in one MZI at a time. For each MZI, we perform 1000 Monte Carlo iterations and calculate the average $RVD$. In each iteration, the MZI parameters ($\theta$,~$\phi$,~$r$,~$r'$,~$t$,~$t'$) corresponding to the faulty MZI are chosen from a Gaussian distribution with $\sigma_{PhS}=\sigma_{BeS}=$~0.05. From Fig. \ref{U1234} we observe that there is a significant variation in the average $RVD$ corresponding to different MZIs representing the same unitary matrix. Note also that the distribution of average $RVD$ across the MZIs differs across the four unitary matrices. Thus, it is clear that the impact of uncertainties in the MZI array on the accuracy of the unitary multiplier varies from case to case.

\subsection{System-Level: SPNN}
Variations in the MZI parameters lead to faulty matrix multiplications in the linear layers, imposing classification accuracy loss in SPNNs. To show the severe impact of such variations in SPNNs, we present a case study of an SPNN handling the MNIST hand-written digit classification task \cite{lecun1998mnist}. \par
To convert the 28$\times$28~$=$~784 dimensional real-valued images in the MNIST dataset to complex-valued vectors, we consider the shifted fast Fourier transform of each image; this results in a 784-dimensional complex-valued vector for each image. To compress the feature vector, we consider the values within 4$\times$4 region at the center of the frequency spectrum. Compared to the baseline accuracy of 94.12\% with the 28$\times$28 feature vector, the 4$\times$4 case results in only 6.77\% accuracy loss. \par

In our SPNN architecture, fully connected feedforward networks with two hidden layers of 16-complex valued neurons are implemented using the Clements design \cite{clements2016optimal}. Each linear layer is followed by the nonlinear Softplus function applied to the modulus of the complex numbers. To model intensity measurement, a modulus squared nonlinearity is applied after the output layer. This is followed by a final LogSoftMax layer to obtain a probability distribution. We use a cross-entropy loss function during training \cite{cover2006elements}. \par

We realize the three weight matrices corresponding to the neurons in the input and the two hidden layers in our SPNN using MZI arrays. Based on our network architecture, the dimensions of the weight matrices are 16$\times$16 (input layer), 16$\times$16 (first hidden layer), and 16$\times$10 (second hidden layer). To analyze the impact of random uncertainties in the MZI arrays on the SPNN, we perform the following experiments:
\begin{itemize}[leftmargin=*]
    \item $EXP_1$ (global uncertainties): We select a $\sigma_{PhS}$ and $\sigma_{BeS}$ and for each selected value, perform 1000 Monte Carlo iterations. For each iteration, we calculate the inferencing accuracy using the 10000 test images in the MNIST dataset. The use of 1000 Monte Carlo iterations is formally justified based on the fact that with a 95\% confidence interval, the maximum margin of error in the mean of the inferencing accuracy is 6.27\%, which is within the acceptable range \cite{acceptablemargin}. Note that $EXP_1$ is performed with uncertainties inserted only in PhS, only in BeS, and in both where $\sigma_{PhS}=\sigma_{BeS}$.  
    \item $EXP_2$ (global uncertainties with zonal perturbations): To find the impact of localized uncertainties on the SPNN accuracy, we divide the SPNN into different zones, each consisting of four MZIs arranged in a 2$\times$2 grid. We insert random perturbations with $\sigma_{PhS}=\sigma_{BeS}=$~0.1  in a selected zone while the remaining zones have uncertainties with $\sigma_{PhS}=\sigma_{BeS}=$~0.05. For each selected zone, we again consider 1000 Monte Carlo iterations (similar to $EXP_1$) and calculate the reduction in the mean inferencing accuracy from the nominal case.   
\end{itemize}
\begin{figure}[t]
  \centering
  \includegraphics[scale=0.49]{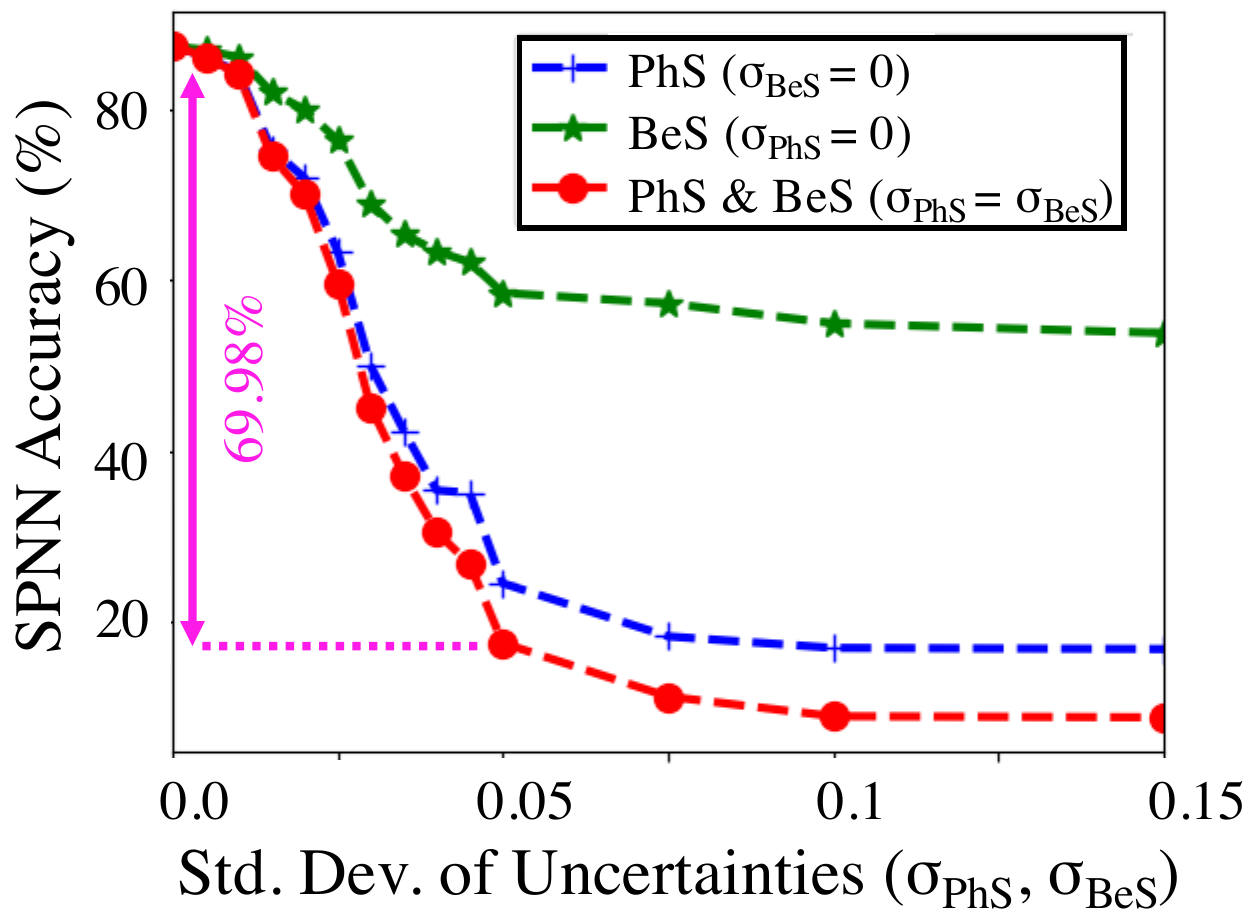}
  \caption{Impact of random uncertainties in the SPNN components (PhS and BeS) on the SPNN inferencing accuracy ($EXP_1$).}
  \vspace{-1.6em}
  \label{compress_EXP1}
\end{figure}

Fig. \ref{compress_EXP1} shows the simulation results for $EXP_1$ when uncertainties are inserted in (i) only PhS, (ii) only BeS, and (iii) both PhS and BeS. For all these cases, the accuracy declines steeply as $\sigma$ increases before it saturates around $\sigma=$~0.075 where the accuracy drops below 10\% (accuracy associated with a random guess). Also, we can see that uncertainties in PhS have a higher impact on accuracy compared to those in BeS. \par

\begin{figure}[t]
  \centering
  \includegraphics[scale=0.49]{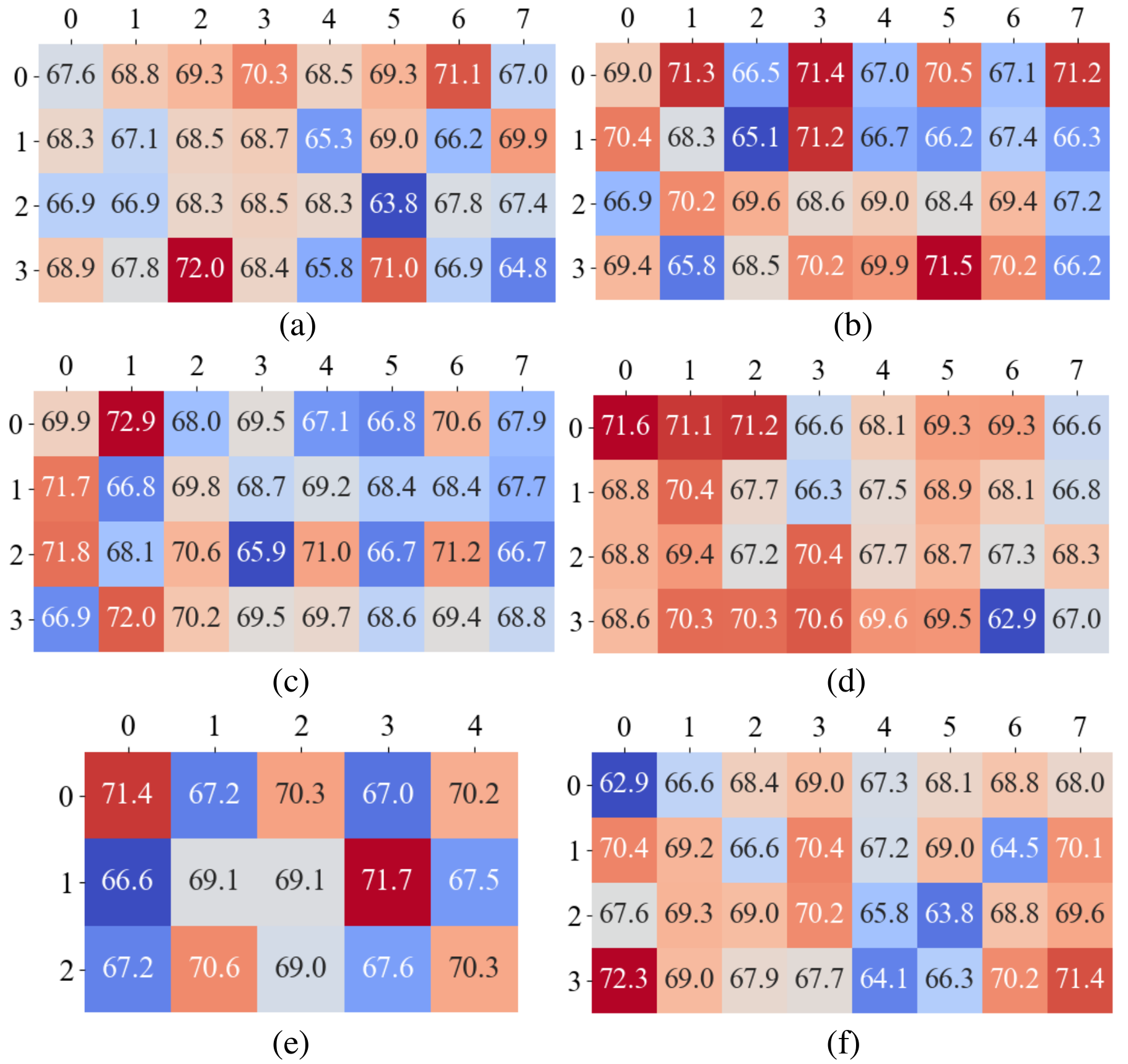}
  \caption{Accuracy loss (\%) due to zonal perturbations in linear layers ($EXP_2$): (a) $U_{L0}$, (b) $V^H_{L0}$, (c) $U_{L1}$, (d) $V^H_{L1}$, (e) $U_{L2}$, and (f) $V^H_{L2}$.}
  \vspace{-1.7em}
  \label{heatmap}
\end{figure}

The three linear layers in our SPNN can be represented by six unitary multipliers. The impact of zonal perturbations in these unitary multipliers on the classification accuracy (experiment $EXP_2$) is presented as heatmaps in Fig.~\ref{heatmap}. Figs. \ref{heatmap}(a)--(b) correspond to the $U$ and $V^H$ matrices of the first linear layer while Figs.~5(c)--(d) and Figs.~5(e)--(f) correspond to the second and third linear layers, respectively. Note that for all these cases, the diagonal matrix $\Sigma$ is assumed to be error-free with the singular values arranged in random order. Each box in the heatmaps corresponds to a zone with the height (width) of the layer increasing vertically (horizontally). The value (color) in each box signifies the accuracy loss when a zonal perturbation is applied to the corresponding zone. From experiment $EXP_1$ (Fig. \ref{compress_EXP1}), we know that the reduction in SPNN accuracy under a global uncertainty of $\sigma_{PhS}=\sigma_{BeS}=$~0.05 is 69.98\%. Fig.~\ref{heatmap} shows that even under zonal perturbations, the accuracy loss hovers around 69.98\%. However, in some zones, the zonal perturbations result in a decreased accuracy loss (e.g., the zone in row 2 column 5 in Fig. \ref{heatmap}(a)), whereas in others they exacerbate the impact of global uncertainties (e.g., the zone in row 3 column 0 in Fig.~\ref{heatmap}(f)). Moreover, note that the low- and high-impact zones are arranged randomly in each unitary multiplier. This shows that the impact of localized uncertainties in MZIs can differ significantly and some MZIs are more critical than others (see also Fig. \ref{U1234}). 

%\vspace{-1em}
\section{Conclusion}
We have modeled the impact of random uncertainties in SPNNs that arise due to fabrication-process variations and thermal crosstalk. Simulation results from our hierarchical approach show that even minor uncertainties in SPNN building blocks have a significant impact on the inferencing accuracy and reliability in SPNNs. Such impact depends on both the tuned parameter values and the position of affected components. The proposed modeling framework can be used to identify and compensate for critical components in SPNNs during design.

\bibliographystyle{IEEEtran}
%\bibliography{Refs}
{\footnotesize
\bibliography{Refs}}
\end{document}